\documentclass{mem}
\usepackage{natbib}\usepackage{txfonts}\usepackage{balance}
\usepackage{graphicx}
\usepackage[a4paper,breaklinks,dvipdfm]{hyperref}
\idline{75}{1}
\begin{document}
\def\teff{$T\rm_{eff }$}
\def\kms{$\mathrm {km s}^{-1}$}

\title{Protostellar accretion and the cosmological lithium problem}

 \subtitle{}

\author{Santi Cassisi\inst{1,2}
          \and
          Maurizio Salaris\inst{3}
         \and
          Scilla Degl'Innocenti\inst{2,4}
          \and
          Pier Giorgio Prada Moroni\inst{2,4}
          \and
          Emanuele Tognelli\inst{2,4}
          }
\institute{INAF - Osservatorio Astronomico d'Abruzzo, Via M. Maggini, I-64100, Teramo, Italy\\
             \and
              INFN - Sezione di Pisa, Largo Bruno Pontecorvo 3, I-56127, Pisa, Italy\\
	     \and
              Astrophysics Research Institute, Liverpool John Moores University, 146 Brownlow Hill, Liverpool L3 5RF, UK
             \and
               Dipartimento di Fisica, Universit\`a di Pisa, Largo Bruno Pontecorvo 3, I-56127, Pisa, Italy\\
          }
     
\authorrunning{S. Cassisi et al.}

\titlerunning{Protostellar accretion in low-mass, metal-poor stars}

\abstract{The cosmological lithium problem, i.e. the discrepancy between the lithium abundance predicted by the Big Bang Nucleosynthesis and the one observed for the stars of the "Spite plateau", is one of the long standing problems of modern astrophysics. A possible astrophysical solution involves lithium burning due to protostellar mass accretion on Spite plateau stars. In present work, for the first time, we investigate with accurate evolutionary computations the impact of accretion on the lithium evolution in the metal-poor regime, that relevant for stars in the Spite plateau.
}
\maketitle{}

\section{Introduction}

The Big Bang Nucleosynthesis (BBN) is one of the underpinnings of the standard Big Bang cosmological model \citep[see,][and references therein]{coc17}. Although, a fine agreement does exist between BBN calculations and observational estimates for the primordial abundances of deuterium and helium, a significant disagreement does exist as far as it concerns the primordial $^7 Li$ abundance. In fact, estimates of the cosmological Li abundance are based on measurements on the so-called Spite plateau \citep[see, e.g.][and refereces therein]{melendez10}: metal-poor, field dwarf stars with ${T_{eff}}$ larger than about 5900~K show a quite uniform surface Li abundance; on the basis of considerations related to Galactic chemical evolution models as well as canonical (non-diffusive) stellar evolutionary predictions, the observed Spite plateau Li abundance should be equivalent to the primordial Li abundance. However, as it is well known, BBN predictions are about 0.3-0.6~dex larger than the Spite plateau value, and this is the long-standing \lq{cosmological Li problem}\rq. Indeed, the situation becomes still more complicated when accounting for the occurrence of diffusive processes in real stars. 

Along these decades several avenues have been suggested to try to solve or at least alleviate the  \lq{cosmological Li problem}\rq; however we still lack a robust scenario for removing this discrepancy between stellar evolution predictions and cosmological prescriptions.

In present work we wish to explore the possible impact on the Pre-Main Sequence/Main Sequence (PMS/MS) Li abundances of the accretion phase  during the protostellar evolution, that is usually not included in PMS calculations. A previous analysis for low-mass, solar metallicity stars \citep{baraffe10} has shown that the accretion phase could indeed alter the starting Li abundance at the beginning of the PMS/MS, reducing it below the initial value at the beginning of accretion. Our purpose is to extend this analysis in the regime of  metal poor low-mass models, to check if accretion during the protostellar phase can actually help to remove the discrepancy between Spite plateau and BBN abundances. 

\section{The evolutionary framework}

The details about the evolutionary code and physical inputs adopted in present analysis can be found in \cite{tognelli20}. Here we briefly note that we have computed the detailed evolution from the protostellar stage up to the MS of models with final masses equal to $0.7$ and $0.8M_\odot$, for metallicity equal to $[Fe/H]\approx-2.1$, values selected as appropriate for the stars in the Spite plateau. When managing the accretion processes during the protostellar phase one has to make some assumptions about some crucial parameters \citep[see][for more details and useful references]{tognelli20}, that are:
\begin{itemize}

\item{{\sl The mass and radius of the seed}: we adopted two values for $M_{seed}$: 1 and 10 Jupiter mass ($M_J$), and  three distinct values for $R_{seed}$: 0.5, 1.5 and $3R_\odot$;}

\item{{\sl accretion rate}: we explored in detail the case of an accretion rate equal to $10^{-5}M_\odot/yr$, but also considered the case of $10^{-6}M_\odot/yr$ and $5\times10^{-5}M_\odot/yr$; in addition to a constant accretion rate, the case of a bursty accretion process has been also investigated;}

\item{{\sl cold versus hot scenario}: when simulating the protostellar accretion process one can assume that no energy is transferred from the accreted matter to the accreting star, that is the {\sl cold} accretion scenario; or alternatively one can assume that some fraction of the energy carried by the accreted matter is added to the internal energy of the star, the case the whole amount of energy of the accreted matter is transferred to the star is named the {\sl hot} accretion scenario; in present investigation we explored both the cases of cold and hot accretion scenario, as well as a mixed case of hot plus cold protostellar accretion.}

\end{itemize}

\begin{figure}[]
\resizebox{\hsize}{!}{\includegraphics[clip=true]{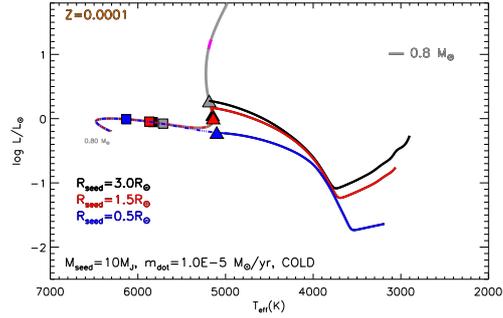}}
\caption{The evolution in the H-R diagram of models computed by using various values for the seed radius, but a fixed value for the seed mass and accretion rate  (see labels), accounting for the cold accretion scenario. In all cases the final stellar mass is $0.8M_\odot$. The gray line corresponds to the evolutionary path of a standard, non accreting model. The location of models at 1~Myr (triangles) and 10~Myr (squares) and the region corresponding to the deuterium burning phase (thick magenta line) are also marked.}
\label{figrseed}
\end{figure}

\begin{figure}[]
\resizebox{\hsize}{!}{\includegraphics[clip=true]{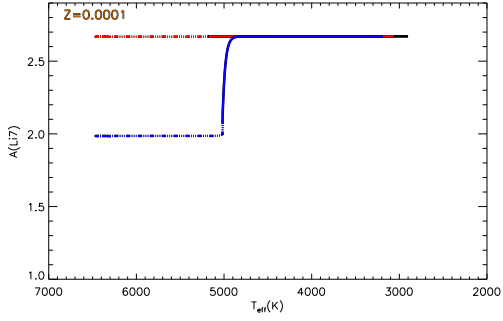}}
\caption{As fig.~\ref{figrseed} but for the trend of the surface Li abundance as a function of the $T_{eff}$. }
\label{figlirseed}
\end{figure}

\section{Results}

\begin{figure}[]
\resizebox{\hsize}{!}{\includegraphics[clip=true]{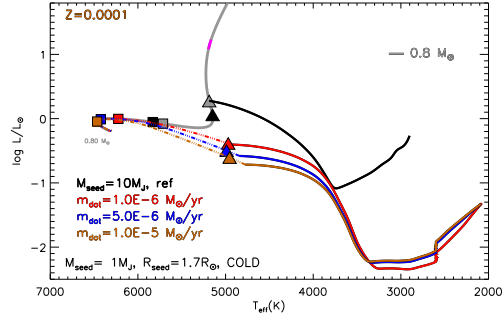}}
\caption{As fig.~\ref{figrseed} but for a different assumption about the mass of the seed. The reference models (black line) corresponds to the model computed by adopting $M_{seed}=10M_J$ and accretion rate $10^{-5}M_\odot/yr$.}
\label{figmseed}
\end{figure}

\begin{figure}[]
\resizebox{\hsize}{!}{\includegraphics[clip=true]{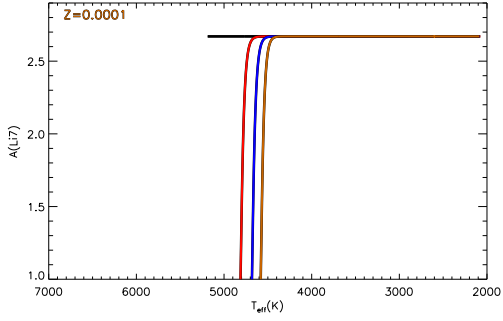}}
\caption{As fig.~\ref{figlirseed} but for the models shown in fig.~\ref{figmseed}.}
\label{figlimseed}
\end{figure}

\begin{figure}[]
\resizebox{\hsize}{!}{\includegraphics[clip=true]{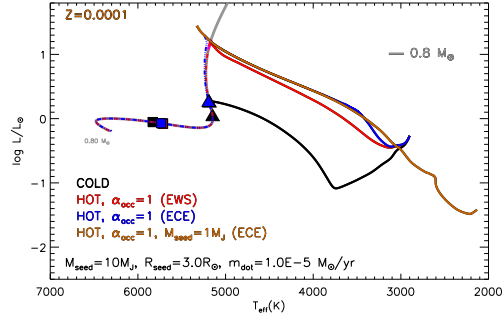}}
\caption{The evolution in the H-R diagram of \emph{hot} accreting models, for a final mass of $0.8M_\odot$, and adopting the labelled values for the seed mass/radius, and accretion rate. The labels \lq{EWS}\rq\ and \lq{ECE}\rq\ correspond to the case when the energy associated to the accreted matter is distributed (uniformly) to the whole stellar structure, or only the outer convection zone, respectively. The black line corresponds to the same accreting models but for the \emph{cold} scenario. The dark yellow line shows the evolution of the same \emph{hot} accreting model but for a seed mass equal to $1M_J$.}
\label{fighot}
\end{figure}

The detailed evolutionary computations performed for this analysis have clearly shown that whenever the accretion process produces bright and extended configuration during the protostellar stage, the resulting structure at the end of the accretion process closely resembles a standard, non-accreting star with the same final mass; this is true concerning both the structural properties and the surface Li abundance. We have verified that the final outcomes of these evolutionary computations hugely depend on some of the adopted initial conditions. More in detail, present results can be summarised as follows:

\begin{itemize}

\item{in the case of \emph{cold accretion}, the PMS evolution and the surface Li abundance are largely affected when adopting small values for $M_{seed}$ and $R_{seed}$ as shown by data shown in figures~\ref{figrseed} to \ref{figlimseed}. One can note that for $M_{seed}=1M_J$ and $R_{seed}$ less than $1R_\odot$; PMS models do not follow the standard Hayashi track. Significant Li depletion can be actually be obtained when using small values for the seed mass and radius, but the results depend strongly on the initial conditions as well as also on the final stellar mass; for instance fig.~\ref{figlirseed} shows that for $R_{seed}=0.5R_\odot$ a significant Li depletion can be obtained for a final mass of $0.8M_\odot$, but our computations reveals that for a final mass of $0.7M_\odot$ the Li depletion is negligible. Data in fig.~\ref{figlimseed} show that for $M_{seed}=1M_J$ a complete surface Li destruction is also possible. Interestingly, the final results are barely affected - if any - by the adopted accretion rate;}

\item{for the \emph{hot accretion} scenario, regardless of the exact choice about the seed mass and radius, the models are barely affected by the accretion process. The reason for this occurrence is due to the fact that hot accreting models are in a more expanded and brighter configuration during the whole protostellar stage in comparison with \emph{cool} accreting models as shown by data in fig.~\ref{fighot}. The same result is obtained for the mixed \emph{hot plus cold protostellar accretion} case. It is also worth noting that, at odds, with standard stellar evolution and \emph{cold} models, the \emph{hot} accretion models do develop a radiative core during the PMS; this is a consequence of the ingestion of the additional energy source associated to the accreted matter;}

\end{itemize}

The physical reasons driving the discussed behaviour for both \emph{cold} and \emph{hot} accreting models have to be searched in the impact that the various assumptions about the initial conditions ($M_{seed}$ and $R_{seed}$) have on the Kelvin-Helmholtz timescale ($\tau_{KH}$), and how this characteristic timescale compares with the accretion time ($\tau_{acc}\propto{1/\dot{m}}$): if $\tau_{KH}<\tau_{acc}$ the PMS evolution is driven by the surface radiative losses, and hence the impact of the accretion process is negligible, on the other hand when  $\tau_{KH}>\tau_{acc}$ the evolution is mainly governed by the accretion process. Since large values for $R_{seed}$ and $M_{seed}$ imply a smaller initial $\tau_{KH}$ value, this explains the previously discussed results.

When considering the evolution of the surface Li abundance, what is really important is the value of temperature at the bottom of the convective envelope and its evolution with the time: in order to have a significant Li depletion it is mandatory not only that this temperature overcomes the typical temperature needed for Li-burning but it has to be larger than this threshold for a significant fraction of the PMS lifetime. Actually these conditions are achieved only in the \emph{cold} models and only when small values for the seed mass and radius are adopted.

\section{Conclusions}

Our results show that a significant reduction of the surface lithium abundance can be obtained only for a restricted range of accretion parameters. However, one has to note that those models, showing a relevant Li depletion, follow a PMS evolution in the HR diagram that is quite different to the one observed for high metallicity PMS stars. The lack of observational data for Galactic, very young and metal-poor, stellar systems, i.e. for PMS stars with Spite plateau metallicities, hampers the possibility to restrict the range of valid accretion parameters and reach firm conclusions. 

This notwithstanding, this first exploratory analysis shows that a fine tuning of the protostellar accretion parameters is required to burn the exact amount lithium during the protostellar/PMS phase --starting from BBN abundances-- to produce the observed constant abundance for the mass and metallicity ranges typical of the Spite plateau.
 
\begin{acknowledgements}
We acknowledge financial support from INFN (iniziativa specifica TAsP) and by PRA (Universit\'a di Pisa 2018-2019, Le stelle come laboratori cosmici di Fisica fondamentale). SC acknowledges support from Premiale INAF MITiC, and  grant AYA2013-42781P from the Ministry of Economy and Competitiveness of Spain.

\end{acknowledgements}

\bibliographystyle{aa}

\end{document}